\documentclass[11pt,a4paper]{article}

\usepackage[utf8]{inputenc}
\usepackage[LGR,T1]{fontenc}
\usepackage{textalpha}
\usepackage{amsmath,amssymb}
\usepackage{graphicx}
\usepackage{booktabs}
\usepackage{hyperref}
\usepackage{url}
\usepackage{natbib}
\usepackage[margin=2.5cm]{geometry}
\usepackage{tabularx}
\usepackage{array}
\usepackage{caption}
\usepackage{float}
\usepackage{microtype}
\usepackage{authblk}
\usepackage{xcolor}

\usepackage{doi}

\hypersetup{
    colorlinks=true,
    linkcolor=blue,
    citecolor=blue,
    urlcolor=blue
}
\title{Tool-calling retrieval versus vector RAG for a small Greek--English knowledge base: accuracy and robustness to how users type Greek}

\author[1]{Nikolaos D.~Tantaroudas\thanks{Corresponding author, \texttt{nikolaos.tantaroudas@iccs.gr}}}
\author[2]{Ilias Karachalios}
\author[3]{Andrew J.~McCracken}

\affil[1]{Institute of Communication and Computer Systems (ICCS), National Technical University of Athens, 9 Iroon Polytechniou Str., 15773 Zografou, Athens, Greece}
\affil[2]{University of Thessaly, 413~34 Larisa, Greece}
\affil[3]{DASKALOS-APPS, 183 Rue de l'Abb\'{e} Griffon, 01960, P\'{e}ronnas, France}

\date{}

\begin{document}

\maketitle

\begin{abstract}
Assistants grounded in a small, frequently edited knowledge base can retrieve through tool calls to a live data interface or through vector retrieval-augmented generation (RAG). We compare the two on KyGround, a benchmark of 198 questions drawn from the published records of a Greek--English agricultural platform on Kythera, Greece, with answers verified automatically against the records and each question posed in up to nine forms, including Greek without accents, in capitals and in three Latin-script (Greeklish) schemes. With Claude Haiku 4.5 as router and answer model, a reconstruction of the platform's tool agent answered 71.6\% of canonical Greek questions correctly and vector RAG 95.3\% (difference $-23.6$ percentage points, 95\% CI $-33.1$ to $-15.1$). Letting the router write the vector query changed nothing, and placing the whole knowledge base of about 26,000 tokens in the prompt reached 99.3\%. The tool agent's losses arose in retrieval. Its literal searches returned nothing when the router's arguments did not occur verbatim in a record, for example when it transliterated Greek into Latin script or combined words that occur in a record but not as one phrase, and the agent then abstained. Unaccented and capitalised questions cost the tool agent about 20 points and vector RAG at most 2; accent-insensitive search removed this loss, and matching stemmed tokens raised the tool agent to 83.8\% on canonical Greek. Greeklish cost both designs about 21 to 32 points. Tool interfaces for community knowledge bases need search that tolerates how users type.
\end{abstract}

\noindent\textbf{Keywords.} Retrieval-Augmented Generation, Tool Calling, Model Context Protocol, Greek, Greeklish, Multilingual Robustness, Agricultural Advisory Services, Benchmark

\medskip
\noindent\textit{This paper reports part of the KyGround study; the full report, with all methods and results, has been submitted to Open Research Europe.}

\section{Introduction}
\label{sec:introduction}

Smallholder farmers on remote islands need advice that is local. Which varieties grow on the island, when the harvest starts there, which producer sells what. Large language models (LLMs) can widen access to such advice, provided that their answers are anchored in authoritative local information rather than in the model's parametric memory \citep{tzachor2023,high2026}, and deployed agricultural assistants therefore combine LLMs with retrieval over curated content \citep{singh2024farmer,collis2025}. Most of the producers on Kythera and Antikythera who tried the deployed assistant whose design is studied here found it helpful, but fewer than a quarter would trust its answers without human verification \citep{karachalios2026}.

There are two common ways of grounding an assistant in a knowledge base. Vector retrieval-augmented generation (RAG) splits documents into chunks, embeds and indexes them, and inserts the chunks nearest to the question into the prompt \citep{lewis2020,karpukhin2020}. Tool-calling retrieval lets the model choose among functions that query a live data interface and write their arguments \citep{yao2023,schick2023}, and the Model Context Protocol (MCP) standardises how such tools are offered to LLM applications \citep{mcp2025}. The assistant \textit{Falco eleonorae}, deployed for the farmers and cooperatives of Kythera, Greece, uses tool calls. Its system article \citep{tantaroudas2026ore}, of which an earlier version is available as a preprint \citep{tantaroudas2026}, argues that for a small, curated and frequently edited knowledge base, exact queries to a read-only data interface are more faithful than nearest-neighbour search. The counter-argument concerns how users type. A tool agent depends on a router that picks the right tool and writes arguments that the interface's matching accepts, which is fragile when Greek is typed without accents, in capitals or in Latin characters (``Greeklish'') \citep{toumazatos2024,luo2026}, whereas multilingual embeddings, which place translations and paraphrases close together \citep{chen2024m3}, may absorb such variation.

This paper tests the accuracy claim, under a pre-specified protocol and with one model throughout. Its contributions are:
\begin{itemize}
\item KyGround, a benchmark of 198 questions drawn from the platform's published records, with gold answers verified automatically against the records, each question posed in up to nine surface forms and scored by a deterministic, Greek-aware scorer;
\item a comparison of six conditions that share one router and answer model, including two controls that locate the difference: vector RAG queried through the same router model and function-calling mechanism as the tool agent, and the whole knowledge base in the prompt;
\item an ablation of the data interface's free-text search normalisation, which shows how much of the tool agent's deficit can be engineered away.
\end{itemize}

\section{Related work}
\label{sec:related}

\textbf{Retrieval and long contexts.} RAG couples a generator with a retriever over a non-parametric memory \citep{lewis2020}, usually dense passage retrieval \citep{karpukhin2020}; \citet{wu2026survey} review its components and variants. Multilingual embedding models, such as the M3-Embedding model used here \citep{chen2024m3}, place translations and paraphrases close together. When a corpus fits in the context window, long-context prompting outperformed RAG on average when resources allowed \citep{li2024lc}, although models use information in the middle of a long context less reliably \citep{liu2024lost}.

\textbf{Tool calling.} ReAct interleaves reasoning with actions such as API calls \citep{yao2023}, Toolformer learns when and how to call tools \citep{schick2023}, and the Berkeley Function Calling Leaderboard evaluates function calling from single calls to agentic use \citep{patil2025}. Because the router selects tools from their descriptions, description quality matters \citep{hasan2026}, and with multilingual input, models often fill tool parameters with values in the user's language, which breaks interfaces that expect a fixed one \citep{luo2026}. Generating executable queries rather than retrieving passages suits structured information \citep{hontan2026}.

\textbf{Greek and Greeklish.} Greek users often type without accents, in capitals or in Latin characters. Greeklish has no single mapping, so one word can be written in many ways, and even native speakers can find it hard to read \citep{toumazatos2024}. The formal transliteration is ISO 843 \citep{iso843}, but informal phonetic and visual schemes dominate in messages. Open Greek LLMs \citep{voukoutis2024,roussis2025} and Greek retrieval benchmarks \citep{beta2026} are emerging, but we know of no head-to-head comparison of tool-calling and vector retrieval in Greek, with the orthographic variation that Greek users produce.

\section{Methods}
\label{sec:methods}

\subsection{Knowledge base and benchmark}

The DASK platform serves the Kythera pilot of the PoliRURAL Plus project \citep{poliruralplus}. Its editors maintain bilingual, Greek-primary records of crops, seasonal-calendar entries, glossary entries (including dialect terms), products, agritourism experiences, training modules, articles and frequently asked questions, and a read-only interface serves the published records with server-side editorial gating \citep{tantaroudas2026ore}. The knowledge base is a frozen snapshot of the platform's public pages including 103 published records, or 78,088 characters (about 26,000 tokens) when serialised bilingually. Every condition answered from in-memory copies of it.

KyGround release contains 198 items including 121 generated from 19 bilingual templates over typed fields (for example the harvest period of a crop or the duration of an experience) and 77 written individually, including aggregations across records. Fourteen items ask for information that the knowledge base does not contain, and twelve for a price that is shown only on request. Each answer is a set of typed facts (text, list members, numbers, ranges, durations, months, price on request, abstention), and an answer is correct when it contains every required fact. Every item was verified automatically against the snapshot, without a model in the loop. Its gold records exist and are published, template items regenerate exactly, every gold fact and every quoted piece of evidence occurs in the records, and aggregates recompute to the same records. Items about the same record share a cluster. A development split of 40 items, made of whole groups of connected records, was used only to tune vector RAG, and the primary analysis set consists of the 148 answerable test items in 79 clusters. Each item was posed in canonical Greek, English, Greek without accents, Greek in capitals, three Greeklish schemes (a transcription in the style of ELOT 743 and ISO 843, an informal phonetic scheme and an informal visual scheme that uses digits and look-alike letters), other inflection and colloquial or regional wording (Table~\ref{tab:forms}). Other-inflection paraphrases exist for 8 items and colloquial or regional variants for 75, so the release contains 1,469 queries.

\begin{table}[htbp]
\centering
\caption{Surface forms of one benchmark item, the harvest period of the olive on Kythera (gold answer: November to January).}
\label{tab:forms}
\small
\begin{tabularx}{\textwidth}{@{}l X@{}}
\toprule
\textbf{Form} & \textbf{Query} \\
\midrule
Canonical Greek & Ποια είναι η περίοδος συγκομιδής για την καλλιέργεια «Ελιά» στα Κύθηρα; \\
English & When is the harvest period for ``Olive'' on Kythera? \\
Greek, no accents & Ποια ειναι η περιοδος συγκομιδης για την καλλιεργεια «Ελια» στα Κυθηρα; \\
Greek, capitals & ΠΟΙΑ ΕΙΝΑΙ Η ΠΕΡΙΟΔΟΣ ΣΥΓΚΟΜΙΔΗΣ ΓΙΑ ΤΗΝ ΚΑΛΛΙΕΡΓΕΙΑ "ΕΛΙΑ" ΣΤΑ ΚΥΘΗΡΑ; \\
Greeklish, ELOT/ISO style & Poia einai i periodos sygkomidis gia tin kalliergeia "Elia" sta Kythira; \\
Greeklish, phonetic & pia ine i periodos sigkomidis gia tin kalliergia "elia" sta kithira; \\
Greeklish, visual & poia einai h periodos sugkomidhs gia thn kalliergeia "elia" sta ku8hra; \\
Greek, other inflection & Πότε μαζεύονται οι ελιές στα Κύθηρα; \\
Greek, colloquial & Ποια είναι η περίοδος συγκομιδής για την καλλιέργεια «Ελιά» στο Τσιρίγο; \\
\bottomrule
\end{tabularx}
\end{table}

\subsection{Conditions}

All six conditions (Table~\ref{tab:conditions}) use Claude Haiku 4.5 \citep{anthropic2025haiku} as router and answer model and share the base prompt and the knowledge base. C2 reconstructs the deployed tool agent. The router selects up to two of nine tools (knowledge search, crop information, seasonal calendar, glossary, products, experiences, cooperatives, training modules and lesson content) and fills their arguments, and an in-process replica of the data interface executes them with its gating, Greek fallback and matching. The interface's free-text matching was characterised before the runs with 50 probes of the platform's public training-module search. It ignored case and matched within words, but it was sensitive to accents, word order and inflection and did not match Greeklish, and the replica was set to the same behaviour (phrase matching with case folding). C1 embeds bilingually serialised chunks of at most 800 characters with BGE-M3 \citep{chen2024m3} and inserts the eight nearest to the question, settings chosen on the development split by the share of gold records retrieved, without model calls. C1q differs from C1 only in that the router model writes the search query, through a single search tool with the same name and query description as the knowledge search of C2, so C1q and C2 share the model, the prompt and the function-calling mechanism and differ in what the call retrieves. C3 gives the answer model both the tool results of C2 and the chunks of C1. C4 places the whole knowledge base in the prompt and shows what the answer model achieves when no evidence can be missed, and C0 answers without retrieval. In an ablation, C2 answered all Greek and Greeklish forms with the replica's free-text search set to (a) phrase matching with case folding, the platform's behaviour, (b) phrase matching with accents also removed and (c) matching of all stemmed tokens.

\begin{table}[htbp]
\centering
\caption{Conditions. All use Claude Haiku 4.5 as router and answer model and share the base prompt, the conversation-history policy and the knowledge-base snapshot.}
\label{tab:conditions}
\small
\begin{tabularx}{\textwidth}{@{}l >{\raggedright\arraybackslash}p{4.2cm} X@{}}
\toprule
\textbf{Code} & \textbf{Condition} & \textbf{Retrieval before the answer} \\
\midrule
C0 & Closed book & none \\
C1 & Vector RAG & the 8 chunks (at most 800 characters) of the bilingually serialised published records nearest to the question (BGE-M3 embeddings, exact cosine search) \\
C1q & Vector RAG, router-formulated query & as C1, for the query that the router model writes when calling a single search tool \\
C2 & Tool retrieval (deployed design, reconstructed) & the router selects up to two of nine tools and fills their arguments. A replica of the data interface executes them \\
C3 & Hybrid & the tool results of C2 and the chunks of C1 \\
C4 & Full context (reference) & the whole published knowledge base in the system prompt \\
\bottomrule
\end{tabularx}
\end{table}

\subsection{Generation}

No model was called through an API. Replies were generated by sub-agents of a Claude agent session run on the authors' subscription, each in its own context window \citep{anthropic2026subagents}. Every request was keyed by the hash of its content and looked up in a cache, and unanswered requests were exported in batches, answered by Claude Haiku 4.5 sub-agents and validated. Check codes placed every 100 lines of a batch, introduced after two early batches of full-context (C4) requests were found to have been read only in part, verified that a sub-agent had read its whole batch. The earlier batches of other requests, whose replies had each been matched to their request, were kept. Because a sub-agent sees every request of its batch, all replies were audited for facts or records that occurred in another request of the batch but not in their own. The 99 flagged replies of the whole study (1.1\% of 9,079 requests) were generated again in isolation, one request per sub-agent, and the results below use these replies. The audit, the withdrawal rule and an isolation control were added after the results had been seen. The withdrawal rule, which covered every flagged reply whatever its grade, was fixed before any regenerated reply was seen. The protocol, hypotheses and analysis plan were fixed, and their digests recorded, before any run with a real model. As the protocol was not deposited in a registry, the study is pre-specified rather than pre-registered.

\subsection{Scoring and statistics}

A deterministic scorer normalises answers and gold facts (Unicode normalisation, case folding, removal of accents and folding of the final sigma), matches them with light stemming and across scripts, and parses numbers, Greek and English month names and ranges, and durations. Abstention and ``price on request'' are recognised by bilingual cue lists. Claude Sonnet 5.5 \citep{anthropic2026sonnet}, a larger model of the same family, served as a second scorer \citep{zheng2023}. It graded all canonical Greek test answers and a sample of 300 English ones, blind to the condition. The deterministic scorer also graded a controlled suite of 1,738 answers whose correctness is known by construction, and the judge a random sample of 350 of them. Accuracies carry 95\% cluster-bootstrap intervals with 10,000 resamples \citep{efron1993}. The pre-specified hypotheses compare the canonical Greek accuracy of C2 with that of C1 (H1), C1q (H2) and C0 (H3), using a two-sided, cluster-level sign-flip permutation test with Holm's correction \citep{holm1979}. H1 and H2 are supported if the adjusted $p < 0.05$, whatever the direction of the difference, and H3 if, in addition, C2 is the more accurate. A robustness gap is the paired difference between a form and canonical Greek, and the difference between the gaps of C2 and C1 is a paired difference-in-differences.

\section{Results}
\label{sec:results}

\subsection{Accuracy on canonical questions}

\begin{table}[htbp]
\centering
\caption{Accuracy on canonical Greek and English questions, \% of the 148 answerable test questions answered fully correctly (95\% cluster-bootstrap CI; 79 clusters).}
\label{tab:canonical}
\small
\begin{tabular}{@{}lll@{}}
\toprule
\textbf{Condition} & \textbf{Greek} & \textbf{English} \\
\midrule
C2 tool retrieval & 71.6 (62.2--79.9) & 79.1 (70.1--86.5) \\
C1 vector RAG & 95.3 (91.3--98.3) & 94.6 (90.6--97.9) \\
C1q vector RAG, router query & 96.6 (93.2--99.3) & 93.9 (89.9--97.3) \\
C3 hybrid & 96.6 (93.2--99.3) & 95.3 (91.6--98.4) \\
C4 full context & 99.3 (97.7--100.0) & 97.3 (94.6--99.4) \\
C0 closed book & 12.2 (7.3--17.8) & 10.1 (5.6--15.4) \\
\bottomrule
\end{tabular}
\end{table}

\begin{figure}[htbp]
    \centering
    \includegraphics[width=0.92\textwidth]{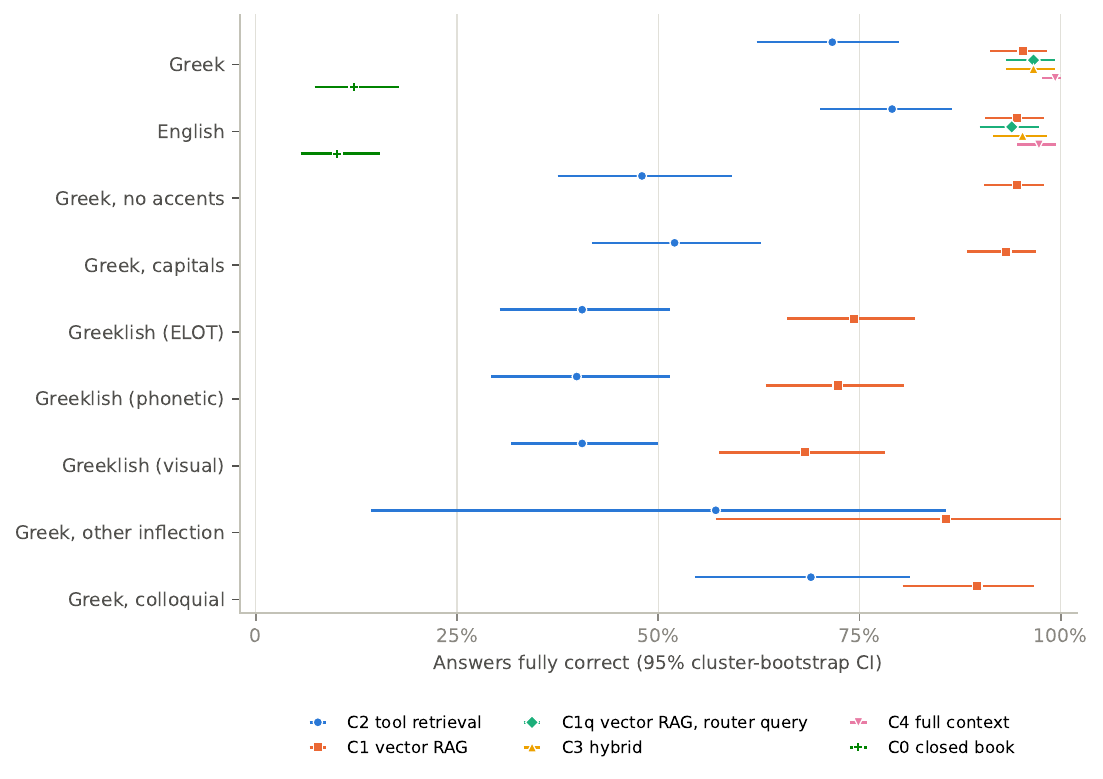}
    \caption{Accuracy by condition and surface form: proportion of answerable test questions answered fully correctly, with 95\% cluster-bootstrap intervals. C1q, C3, C4 and C0 were run on canonical Greek and English only. Conditions are distinguished by marker shape as well as colour; the numbers are in Tables~\ref{tab:canonical} and~\ref{tab:robust}.}
    \label{fig:forms}
\end{figure}

On canonical Greek questions, C2 answered 71.6\% correctly and C1 95.3\% (Table~\ref{tab:canonical}; Figure~\ref{fig:forms}). The paired difference was $-23.6$ percentage points (pp; 95\% CI $-33.1$ to $-15.1$; Holm-adjusted $p < 0.001$), with 37 questions answered correctly by C1 alone and 2 by C2 alone. C2 also trailed C1q by 25.0 pp ($-34.4$ to $-16.6$) and exceeded C0 by 59.5 pp ($+48.5$ to $+69.1$), so all three hypotheses were supported, H1 and H2 in the direction that favours vector retrieval. Letting the router write the vector query did not change accuracy (C1q $-$ C1: $+1.4$ pp; $-1.8$ to $+4.8$; $p = 0.688$), so the gap lies in what the tool calls retrieve rather than in letting a model restate the question. Adding the chunks of C1 to the tool results of C2 (C3) gave 96.6\%, and the whole knowledge base in the prompt (C4) gave 99.3\%. In English, C2 answered 79.1\% and C1 94.6\%.

\subsection{Where the tool agent fails}

\begin{figure}[htbp]
    \centering
    \includegraphics[width=0.85\textwidth]{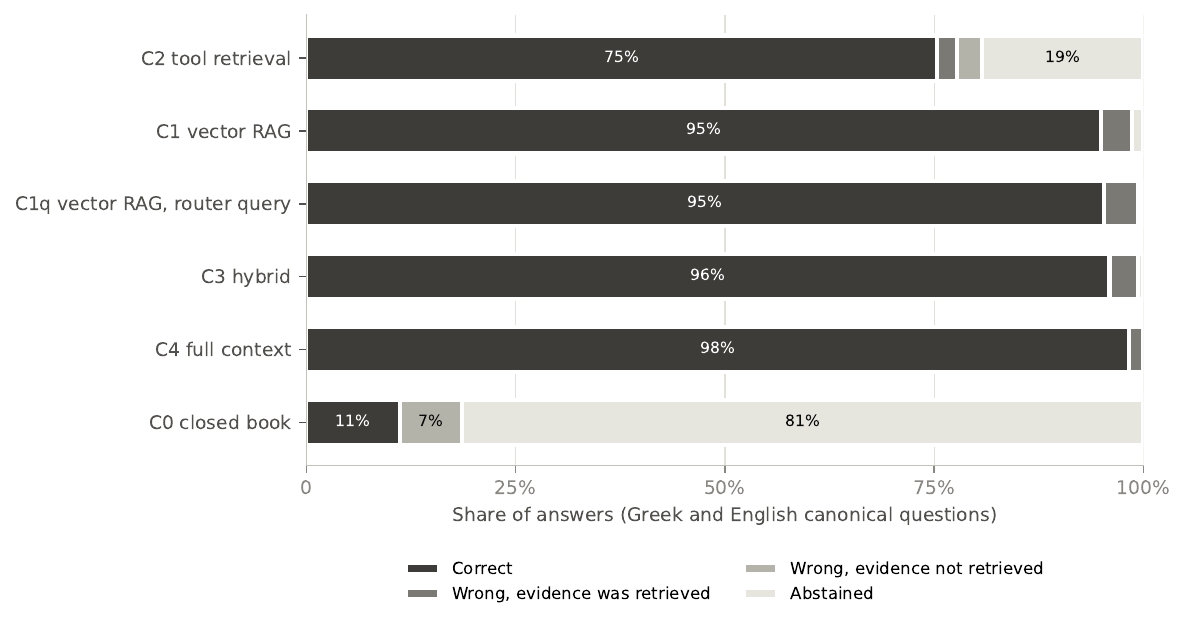}
    \caption{Decomposition of the answers to canonical Greek and English questions (296 per condition): correct; wrong although the gold record had been retrieved; wrong without the gold record among the retrieved records; and abstentions. C0 does not retrieve anything, so its wrong answers count as wrong without retrieved evidence.}
    \label{fig:decomp}
\end{figure}

The gold record was among the retrieved records for 75.7\% of the answers of C2 to canonical Greek and English questions, against 98.6\% for C1 (Figure~\ref{fig:decomp}). Wrong answers despite retrieved evidence were rare in every condition with retrieval (1.7\% to 4.1\% of answers). The shortfall of C2 consisted mainly of abstentions (19.3\% of its answers), in which the agent said that the information was not available after its tool calls had not returned the relevant record. C2 made 1.38 tool calls per question, and in 23.3\% of its answers at least one call returned no record. In an exploratory analysis, its accuracy depended on whether the question quoted a record's exact title. It answered 81.9\% of the 105 canonical Greek questions that did and 46.5\% of the 43 that did not, against 97.1\% and 90.7\% for C1. The router's arguments show why searches failed (an exploratory analysis). Although the search tool asks for terms ``preferably in Greek'', the router wrote 28.7\% of its free-text arguments for canonical Greek questions in Latin script, as transliterations (for example ``Thymari meli Kythiron 1kg'' for «Θυμαρίσιο Μέλι Κυθήρων 1kg») or English keywords, and 46.9\% of these returned no record, against 26.2\% of the arguments written in Greek.

\subsection{Robustness to Greek surface forms}

\begin{table}[htbp]
\centering
\caption{Accuracy of the tool agent (C2) and vector RAG (C1) by surface form, \% (95\% cluster-bootstrap CI), and the difference between their gaps relative to canonical Greek, percentage points (paired difference-in-differences, 95\% CI). Other inflection has 7 test questions and colloquial wording 58; every other form has 148.}
\label{tab:robust}
\small
\begin{tabularx}{\textwidth}{@{}>{\raggedright\arraybackslash}X l l l@{}}
\toprule
\textbf{Form} & \textbf{C2 tool retrieval} & \textbf{C1 vector RAG} & \textbf{Gap difference, C2 $-$ C1} \\
\midrule
Greek & 71.6 (62.2--79.9) & 95.3 (91.3--98.3) & reference \\
English & 79.1 (70.1--86.5) & 94.6 (90.6--97.9) & +8.1 ($-$0.7 to +16.7) \\
Greek, no accents & 48.0 (37.5--59.2) & 94.6 (90.5--97.9) & $-$23.0 ($-$35.1 to $-$10.6) \\
Greek, capitals & 52.0 (41.8--62.8) & 93.2 (88.4--97.0) & $-$17.6 ($-$27.9 to $-$6.0) \\
Greeklish (ELOT) & 40.5 (30.3--51.4) & 74.3 (66.0--81.9) & $-$10.1 ($-$24.6 to +5.3) \\
Greeklish (phonetic) & 39.9 (29.2--51.5) & 72.3 (63.4--80.5) & $-$8.8 ($-$24.5 to +7.9) \\
Greeklish (visual) & 40.5 (31.7--50.0) & 68.2 (57.6--78.2) & $-$4.1 ($-$18.3 to +11.1) \\
Greek, other inflection & 57.1 (14.3--85.7) & 85.7 (57.1--100.0) & +0.0 ($-$42.9 to +42.9) \\
Greek, colloquial & 69.0 (54.5--81.2) & 89.7 (80.4--96.7) & +12.1 (+0.0 to +25.4) \\
\bottomrule
\end{tabularx}
\end{table}

Relative to canonical Greek, unaccented questions lowered the accuracy of C2 by 23.6 pp and that of C1 by 0.7 pp, and capitals by 19.6 and 2.0 pp; the differences between the two systems' gaps were $-23.0$ pp ($-35.1$ to $-10.6$) and $-17.6$ pp ($-27.9$ to $-6.0$) (Table~\ref{tab:robust}). Greeklish lowered the accuracy of both systems, by 31.1 to 31.8 pp for C2 and by 20.9 to 27.0 pp for C1, leaving C1 at 68.2\% to 74.3\% and C2 at 39.9\% to 40.5\%, and the differences between these gaps could not be statistically distinguished. With Greeklish questions the router wrote 61\% to 63\% of its free-text arguments in Latin script, and 74\% to 80\% of these returned no record. Colloquial and regional wording changed the accuracy of C2 by $+5.2$ pp and that of C1 by $-6.9$ pp; the other-inflection paraphrases (7 test items) are too few for inference.

\subsection{Search normalisation}

\begin{table}[htbp]
\centering
\caption{Accuracy of the tool agent (C2) under three free-text search normalisations of the data interface, \% (95\% cluster-bootstrap CI): (a) phrase matching with case folding, the platform's behaviour and the main run; (b) phrase matching with accents also removed; (c) matching of all stemmed tokens.}
\label{tab:ablation}
\small
\begin{tabularx}{\textwidth}{@{}>{\raggedright\arraybackslash}X l l l@{}}
\toprule
\textbf{Form} & \textbf{(a) Phrase, case folded} & \textbf{(b) Accents also folded} & \textbf{(c) Stemmed tokens} \\
\midrule
Greek & 71.6 (62.2--79.9) & 73.6 (64.4--81.9) & 83.8 (77.6--89.2) \\
Greek, no accents & 48.0 (37.5--59.2) & 73.6 (64.5--81.8) & 83.8 (77.8--89.5) \\
Greek, capitals & 52.0 (41.8--62.8) & 73.0 (64.4--81.2) & 81.8 (75.0--88.4) \\
Greeklish (ELOT) & 40.5 (30.3--51.4) & 43.2 (32.7--54.4) & 51.4 (41.0--62.2) \\
Greeklish (phonetic) & 39.9 (29.2--51.5) & 43.9 (33.1--55.3) & 49.3 (38.6--60.5) \\
Greeklish (visual) & 40.5 (31.7--50.0) & 44.6 (35.6--54.0) & 52.0 (43.6--60.8) \\
Greek, other inflection & 57.1 (14.3--85.7) & 57.1 (14.3--85.7) & 71.4 (42.9--100.0) \\
Greek, colloquial & 69.0 (54.5--81.2) & 69.0 (54.5--81.2) & 74.1 (61.1--85.1) \\
\bottomrule
\end{tabularx}
\end{table}

Folding accents in the interface's free-text search (mode b) removed the losses of C2 on unaccented and capitalised Greek and raised its canonical Greek accuracy from 71.6\% to 73.6\% (Table~\ref{tab:ablation}). Matching all stemmed tokens (mode c) raised it to 83.8\%, and on the questions that do not quote a record's title from 46.5\% to 72.1\%. Neither mode helped much with Greeklish (43.2\% to 52.0\%), because transliterated arguments match neither the Greek nor the English text of the records. Even with stemmed tokens, C2 remained 11.5 pp below C1 on canonical Greek.

\subsection{Unanswerable questions, prices and context size}

No condition gave an amount for a price shown only on request (0 of 20 answers each). On unanswerable items, the deterministic scorer counted at most 2 of 20 answers per condition as hallucinations, but all 9 such answers were abstentions in wordings that its cue lists did not cover. Per question, C1 used one model call and 1,612 input tokens on average, C1q two calls and 2,069 tokens, C2 2.98 calls and 6,517 tokens, and C4 one call and 26,294 tokens, estimated with the o200k\_base encoding \citep{tiktoken}.

\subsection{Validation of scoring and of batched generation}

The deterministic scorer graded all 1,738 answers of the controlled suite as constructed, as did the LLM judge on the 350 that it graded. On 1,248 judged test answers, the two scorers agreed on 97.4\% ($\kappa = 0.92$; \citealp{cohen1960}), and with the judge's grades the paired differences of C2 from C1, C1q and C0 were $-25.7$, $-25.7$ and $+60.1$ pp, leading to the same conclusions. Of the 76 accuracy answers that the contamination audit flagged, 57 answered Greeklish questions, for which retrieval most often failed, and 53 had been scored correct. Generated again in isolation, 9 were correct. The correction changed the primary contrasts by less than one point but lowered the Greeklish accuracy of C1 by 5.4 to 8.1 points: with the batched replies, the differences between the Greeklish gaps of C2 and C1 would have been $-14.9$, $-16.2$ and $-9.5$ pp, instead of $-10.1$, $-8.8$ and $-4.1$ pp. In a control, 95 of 100 unflagged replies generated again in isolation received the same grade, 4 of the 5 discordant replies favoured the batched version, so a residual inflation of a few points among unflagged replies cannot be excluded.

\section{Discussion}
\label{sec:discussion}

With one router and answer model, vector RAG answered questions about a small, community-edited knowledge base more accurately than tool retrieval, and the controls locate the difference. C1q gave vector retrieval the same router model, prompt and function-calling mechanism as C2 and changed nothing, and C4 shows that the answer model answers almost every question when the evidence is present, so retrieval was the bottleneck. The tool agent's literal search returned nothing when the router's arguments did not occur verbatim in a record (a transliterated or translated keyword, an accent added to a title stored in capitals, or keywords that occur in a record but not as one phrase), and the agent then abstained. Multilingual input is known to make models fill tool parameters in an unexpected language \citep{luo2026}. Here the router transliterated Greek into Latin script even though the questions and the records were Greek.

Three lessons follow for tool interfaces over small community knowledge bases. First, free-text search should ignore case, accents and the final sigma and match stemmed tokens rather than literal phrases. In the replica, this removed the whole loss on unaccented and capitalised questions and recovered about half of the gap to vector RAG on canonical Greek. Second, arguments should be in the script of the records, and the tool description alone did not ensure it. The search tool asked for terms preferably in Greek, yet the router wrote 28.7\% of its free-text arguments for canonical Greek questions in Latin script. Third, Greeklish remains hard for both designs. The vector RAG reached 68.2\% to 74.3\% on Greeklish questions, 21 to 27 points below its canonical accuracy, against about 40\% for the tool agent, so a vector fallback would raise the tool agent's Greeklish accuracy without solving Greeklish, and transliteration-aware matching remains to be tested. On canonical questions, adding vector chunks to the tool results (C3) matched the accuracy of vector RAG. For a knowledge base of about 26,000 tokens, placing everything in the prompt was the most accurate design, at about four times the input of the tool agent, and the risk of overlooking evidence in the middle of a long context \citep{liu2024lost} did not materialise at this size.

\textbf{Limitations.} The authors built the system evaluated as C2 and the platform from which the benchmark was drawn. To limit bias, the protocol was fixed before the runs, the primary scoring is deterministic and all outputs are available on request, and on accuracy the results run against the system article's claim. Claude Haiku 4.5 replaced the deployed GPT-5-family models \citep{openai2025gpt5}, so the accuracies estimate what the designs achieve with this model rather than the deployed system's accuracy, and another router may transliterate more or less often. C2 is a reconstruction. Seven of the nine tool names, all nine tool descriptions and the full system prompt were reconstructed. Its router saw only the nine local tools, whereas the production router also chooses among the upstream agent's general tools, and the replica's agreement with the deployed interface could not be measured. The judge is from the same model family as the answer model, so a shared preference for the family's phrasing cannot be excluded, serving only as a second scorer. The questions were derived from the records rather than collected from users, and 105 of the 148 primary questions quote a record's exact title, which favours literal matching, so the tool agent's accuracy on users' own wording may be lower than reported. The gold answers were verified against the records rather than by the platform's editors. The knowledge base is small and comes from one platform. Batched generation contaminated 1.1\% of replies, and the audit detects only borrowed facts, records and markers. Generation in a session cannot be repeated token for token, each query was answered once, and the study can detect differences of about 14 points with 80\% power, so the smaller differences between the Greeklish gaps are reported as estimates.

\section{Conclusions}
\label{sec:conclusions}

For a small Greek--English knowledge base, vector RAG answered 95\% of canonical Greek questions correctly and tool retrieval 72\%. The gap arose in literal search, widened with unaccented and capitalised Greek and was partly closed by accent-insensitive, stemmed matching in the data interface, while Greeklish remained hard for both designs. Tool interfaces should be evaluated in the spellings that users actually type.

\section*{Data and code availability}

{\sloppy The KyGround benchmark, with the names of natural persons replaced by role labels, the run outputs and the evaluation harness are available from the authors on request. The aggregated results behind the tables and figures of this paper are provided as ancillary files. The frozen snapshot of the platform's records, the request cache and the batch files given to the sub-agents contain the text of the records, including names of natural persons, and are not deposited. They are available on request from the corresponding author, Nikolaos D.~Tantaroudas (\texttt{nikolaos.tantaroudas@iccs.gr}), to researchers and reviewers who state their name, affiliation and purpose and agree in writing to use the material only to verify or reproduce the results or for other non-commercial research, not to redistribute it, not to attempt to identify or contact the persons named in it, to delete it when their purpose is complete and to cite this work and the datasets. Requests are answered within 30 days. The platform's published records can be viewed at \url{https://dask-platform.onrender.com/}.\par}

\section*{Acknowledgements}

This work was supported by the European Union's Horizon Europe research and innovation programme through the PoliRURAL Plus project (Grant Agreement No.\ 101136910), via its DEVELOP cascade-funding open call. Views and opinions expressed are those of the authors only and do not necessarily reflect those of the European Union or the granting authority. Neither the European Union nor the granting authority can be held responsible for them. The authors thank the editors and producers of the DASK platform, whose published records form the knowledge base studied here, and the PoliRURAL Plus consortium for the Jackdaw agent infrastructure. Large language models are the objects of this study. In addition, the evaluation software, the individually written benchmark items and the phrase lexicon for colloquial wording, the protocol, the analyses and a first draft of the text were prepared with the assistance of Claude Opus 5.5 (Anthropic), which also ran the harness and launched the sub-agents that generated and graded the replies, under the authors' direction. The authors reviewed and revised all code and text and take full responsibility for the content. The authors designed and built the assistant evaluated as C2. NDT is the founder and legal representative of DASKALOS APPS SAS, with which AJM is affiliated.

\end{document}